\documentclass{aa}  
\usepackage{txfonts}

\def\draftversion{0} 
\usepackage{amssymb,latexsym,graphicx,natbib,eufrak,times,amsmath,xspace,ifthen}

\ifthenelse{\equal{\draftversion}{1}}{
  \usepackage{xcolor}
  \newcommand{\sep}[1]{\par\begin{color}[rgb]{0,0.4,0}\begin{center}\hrule\end{center}\end{color}\par} 
  \newcommand{\todo}[1]{\begin{color}{red}\ \ifthenelse{\equal{#1}{}} {$\bullet\bullet\bullet$} {$\bullet$\ #1 $\bullet$}\end{color}} 
  \newcommand{\idea}[1]{\begin{color}[rgb]{0,0.4,0}\textit{#1}\end{color}} 
  \newcommand{\sk}[1]{\begin{color}[rgb]{0.6,0,0.6}#1\end{color}} 
  \newcommand{\toc}{\par\begin{color}[rgb]{0.6,0,0.6}\begin{center}\hrule\vspace{0.5mm}\begingroup\small\let\cleardoublepage\relax\let\clearpage\relax\mytoc\endgroup\vspace{0.5mm}\hrule\end{center}\end{color}\par} 
  \newcommand{\inprep}{\begin{color}[rgb]{0,0.8,0.8}in preparation\end{color}\xspace}
  
  }{
  \newsavebox{\trashcan}

  \newcommand{\sep}[1]{}
  \newcommand{\todo}[1]{}
  \newcommand{\idea}[1]{}
  
  \newcommand{\sk}[1]{}
  \newcommand{\toc}{}
  \newcommand{\inprep}{in preparation\xspace}
  
  }
 \makeatletter\newcommand\mytoc{\@starttoc{toc}}\makeatother 
\long\def\symbolfootnote[#1]#2{\begingroup%
\def\thefootnote{\fnsymbol{footnote}}\footnote[#1]{#2}\endgroup} 

\newcommand{\eqn}[2][]{Equation#1~\ref{eqn:#2}} 
\newcommand{\fig}[2][]{Figure#1~\ref{fig:#2}}

\newcommand{\sect}[2][]{Section#1~\ref{sec:#2}}
\newcommand{\app}[2][]{Appendix#1~\ref{sec:#2}}

\newcommand{\bb}[1]{\ifmmode \mbox{\boldmath $ #1$} \else  \mbox{\boldmath $#1$} \fi}

\newcommand{\mh}{\ensuremath{\textrm{\,--\,}}}    
\newcommand{\U}[1]{\ensuremath{\mathrm{~#1}}}     

\newcommand{\yr}{\U{yr}}
\newcommand{\Myr}{\U{Myr}}          \newcommand{\myr}{\Myr}
\newcommand{\Gyr}{\U{Gyr}}          
\newcommand{\pc}{\U{pc}}
\newcommand{\kpc}{\U{kpc}}
\newcommand{\Mpc}{\U{Mpc}}          
\newcommand{\Msun}{\U{M}_{\odot}}   \newcommand{\msun}{\msunyr}
\newcommand{\Msunyr}{\Msun\yr^{-1}} \newcommand{\msunyr}{\Msunyr}
   
\newcommand{\cc}{\U{cm^{-3}}}
\newcommand{\K}{\U{K}}

\newcommand{\mach}{\ensuremath{\mathcal{M}}}      
\newcommand{\hii}{H{\sc ii} }                     
\newcommand{\htwo}{H$_2$}                          
\newcommand{\aco}{\ensuremath{\alpha_\mathrm{CO}}\xspace}     
\newcommand{\sigg}{\ensuremath{\Sigma_\mathrm{gas}}\xspace}   
\newcommand{\sigsfr}{\ensuremath{\Sigma_\mathrm{SFR}}\xspace} 
\newcommand{\tdep}{\ensuremath{t_\mathrm{dep}}\xspace}        
\newcommand{\jup}{\ensuremath{J_\mathrm{upper}}\xspace}       
\newcommand{\ha}{\ensuremath{\mathrm{H}\alpha}\xspace}       

\newcommand{\saclay}{Laboratoire AIM Paris-Saclay, CEA/IRFU/SAp, Universit\'e Paris Diderot, F-91191 Gif-sur-Yvette Cedex, France}

\newcommand{\toulouse}{Institut de Recherche en Astrophysique et Plan\'etologie (IRAP), CNRS, 14 avenue Belin, 31400 Toulouse, France}

\newcommand{\mpia}{Max-Planck-Institut f\"ur Astronomie/K\"onigstuhl 17 D-69117 Heidelberg, Germany}

\newcommand{\lund}{Department of Astronomy and Theoretical Physics, Lund Observatory, Box 43, SE-221 00 Lund, Sweden}

\newcommand{\roe}{Institute for Astronomy, University of Edinburgh, Royal Observatory, Blackford Hill, Edinburgh EH9 3HJ, UK}

\newcommand{\maryland}{Department of Astronomy, University of Maryland, College Park, Maryland 20742, USA}

\begin{document}
\title{A diversity of starburst-triggering mechanisms in interacting galaxies and their signatures in CO emission}
\titlerunning{A diversity of starburst-triggering mechanisms}

\author{F. Renaud\inst{1}, F. Bournaud\inst{2}, O. Agertz\inst{1}, K. Kraljic\inst{3}, E. Schinnerer\inst{4}, A. Bolatto\inst{5}, E. Daddi\inst{2}, A. Hughes\inst{6}} 
\authorrunning{Renaud et al.}
\institute{$^1$ \lund\\\email{florent@astro.lu.se}\\
$^2$ \saclay\\
$^3$ \roe\\
$^4$ \mpia\\
$^5$ \maryland\\
$^6$ \toulouse}


\abstract{The physical origin of enhanced star formation activity in interacting galaxies remains an open question. Knowing whether starbursts are triggered by an increase of the quantity of dense gas or an increase of the star formation efficiency would improve our understanding of galaxy evolution and allow to transpose the results obtained in the local Universe to high redshift galaxies. In this paper, we analyze a parsec-resolution simulation of an Antennae-like model of interacting galaxies. We find that the interplay of physical processes has complex and important variations in time and space, through different combinations of mechanisms like tides, shear and turbulence. These can have similar imprints on observables like depletion time and CO emission. The densest gas ($>10^4 \cc$) only constitutes the tail of the density distribution of some clouds, but exists in large excess in others. The super-linearity of the star formation rate dependence on gas density implies that this excess translates into a reduction of depletion times, and thus leads to a deviation from the classical star formation regime, visible up to galactic scales. These clouds are found in all parts of the galaxies, but their number density varies from one region to the next, due to different cloud assembly mechanisms. Therefore, the dependence of cloud and star formation-related quantities (like CO flux and depletion time) on the scale at which they are measured also varies across the galaxies. We find that the \aco conversion factor between the CO luminosity and molecular gas mass has even stronger spatial than temporal variations in a system like the Antennae. These results raise a number of cautionary notes for the interpretation of observations of unresolved star-forming regions, but also predict that the diversity of environments for star formation will be better captured by the future generations of instruments.}
\maketitle
\section{Introduction}

In the local Universe, galaxy interactions and mergers are often associated with an enhancement of the star formation activity, particularly visible in (ultra) luminous infrared galaxies (ULIRGs, \citealt{Sanders1996, Scudder2012, Knapen2015}). Although this is now routinely reproduced in simulations \citep[e.g.][]{Karl2010, Teyssier2010, Moreno2013, Renaud2014b}, the underlying physical reasons and the main driver(s) of such starbursts remain uncertain. Understanding the physics of enhanced star formation is key to establish whether and how the known empirical and theoretical relations between star formation and its triggers derived in the local Universe can be transposed to the extreme conditions at high redshift, during the first phases of galaxy formation.

To date, it is not clear whether the bursts of star formation rate (SFR) result from an enhancement of the quantity of dense gas due to a compression from galactic scale mechanisms \citep[e.g.][]{Combes1994, Casasola2004, Kaneko2017}, or from an increase of the efficiency of the formation process itself\footnote{Not to be confused with the dimension-less small-scale star formation efficiency (SFE) describing the stellar mass produced from a given amount of gas, typically at sub-cloud scales ($\lesssim 1 \pc$) and often implemented in sub-grid recipes of star formation. For the sake of clarity, we do not use the term ``efficiency'' to describe the galactic-scale quantity but rather use the depletion time, which is directly its inverse. The small scale SFE and the galactic scale depletion time are not directly related \citep[see][]{Semenov2018}.} (see \citealt{Solomon1988, Gracia2008, Michiyama2016}). The former is often quantified by estimating the molecular gas mass, while the latter relates to the depletion time of the gas reservoir ($\tdep = M_{\rm gas}/{\rm SFR}$). A number of observations of local interacting systems have quantified the variations of these two quantities to determine the main driver of starbursts, often as functions of the interaction parameters. The emergent consensus depicts a mixed influence of both effects, with the increase of dense gas content being the dominant one \citep{Pan2018}. 

In the Schmidt-Kennicutt plane (i.e. the relation between the surface densities of gas and of SFR), an increase of the dense gas mass leads to the expected increase of SFR along the canonical relation ($\sigg \propto \sigsfr^N$, with $N\approx 1.4\mh 2.3$, \citealt{Kennicutt1998, Kennicutt2012, Kravtsov2003, Elmegreen2005, Semenov2018}), i.e. at a roughly uniform depletion time ($\sim 1 \Gyr$, \citealt{Bigiel2010, Saintonge2011, Leroy2013}). Such a high dense gas mass is what is observed in e.g. gas-rich turbulent disk galaxies at redshifts $z \gtrsim 1$ \citep[e.g.][]{Tacconi2010}. However, a decrease of the depletion time as observed, e.g. in LIRGs \citep{Klaas2010, Martinez2012, Nehlig2016} translates into a deviation from this relation. A duality of star formation regimes has been proposed by \citet{Daddi2010b} to distinguish galaxies with long and short depletion times, respectively disks and mergers, but independently of their SFR. To reach this conclusion, these authors had to assume a conversion factor \aco between the CO luminosity and the molecular mass (due to the difficulty to detect \htwo, see \citealt{Bolatto2013}). Uncertainties hinder the choice of this factor \citep{Zhu2003, Gracia2008, Pereira2016}. The value of $0.8 \msun \K^{-1} \U{km}^{-1} \U{s} \pc^{-2}$ adopted by Daddi et al. for the mergers is in sharp contrast with that of the disks (4.3 for the Milky Way) and has been highly debated, since an under-estimated \aco would artificially accentuate or even create the differences between the regimes in depletion times. However, the decrease of \tdep is retrieved in parsec-resolution simulations of interacting galaxies in which the transitions between the disk and the merger regimes (and back) can be resolved, without any assumption on \aco \citep{Renaud2014b, Kraljic2014, Fensch2017, Renaud2018}. Furthermore, by post-processing similar simulations to derive \aco, \citet{Renaud2019} showed that the choice a low \aco for ULIRG systems is likely justified \citep[see also][]{Narayanan2011}, for a few $10 \Myr$ after the peak(s) of the their SFR. The positions of starbursting systems (at least in the local Universe) in the Schmidt-Kennicutt plane thus confirm both an elevation of the dense gas content, and a decrease of the depletion time. (Note that this effect gets hidden in variants of this diagram when \sigg is normalized by the free-fall time, \citealt{Krumholz2012, Federrath2013b, Salim2015}.)

Variations of the depletion time have been reported observationally across galactic disks (e.g. \citealt{Meidt2013, Bigiel2015, Usero2015, Pereira2016, Saito2016, Leroy2017, Tomicic2018}) and along the course of interactions either theoretically \citep[e.g.][]{Renaud2014b}, or observationally by considering statistical samples of galaxies at different evolutionary stages \citep[e.g.][]{Feldmann2012, Violino2018, Pan2018}.

For instance, the evolution of modeled interacting systems in the Schmidt-Kennicutt plane reveals that the successive encounters do not induce the same effects (see e.g. Figure 3 of \citealt{Renaud2014b} for an Antennae-like system, and Figure 11 of \citealt{Renaud2018} for a Cartwheel-like galaxy). Distant passages (usually at an early stage of the interaction, before strong dynamical friction has reduced the orbital energy) trigger a violent increase of the star formation activity over large volumes, while the global surface density of gas remains almost unchanged (measured at a scale of $\sim 100 \pc$). However, closer, penetrating encounters (usually at later stages, moments before the coalescence) yield both an increase of the gas and the star formation rate densities, due to most of the activity being concentrated in the galactic nuclei. Such differences are also visible in the star cluster formation rates, where clusters are preferentially formed in early passages instead of at coalescence \citep{Renaud2015}.

These variations in both space and time relate to the extended star formation, and in particular the off-nuclear activity \citep{Barnes2004, Wang2004, Cullen2006, Smith2008, Hancock2009, Chien2010}, and suggest that different physical mechanisms trigger starburst events, through an increase of the dense gas mass, a decrease of the depletion time, or both, and that their interplay and relative importance vary in space and time, as a function of the properties of the galactic encounter. For instance, the importance of nuclear inflows induced by tidal torques from one galaxy onto the other \citep{Keel1985, Barnes1991} increases with decreasing pericenter distance. Similarly, shocks have a strong impact during close passages, when the dense regions of the ISM overlap \citep{Jog1992}. Conversely, tidal and turbulent compression found in simulations \citep{Renaud2014b} are of gravitational origin and are thus remote effects, occurring even during fly-bys and spanning large volumes. 

Understanding these phenomena, their interplay, and their possible observable signatures is thus an important key to pin down the physics of the physics of star formation, and to determine which quantities should be probed in future observations.

To address this problem with simulations, parsec-resolution is necessary to capture the cold, dense ($\sim 10 \K$, $\gtrsim 10^3 \cc$) phase of the ISM, and the details of its turbulent nature. The high cost of such simulations caps the number of cases that can be studied. Here we use an Antennae-like simulation which encompasses a wide diversity of star-forming environments (nuclei, shocks, extended formation). We explore these spatial variations, attempt to connect them to underlying physical process(es), and seek possible differences of their signatures in the CO emission that could be observed in real galaxies. Such differences could be particularly insightful when numbers of observational diagnostics cannot be resolved and only integrated fluxes are available.

\section{Method}

\subsection{Numerical method}
\label{sec:numerical}

The numerical method followed is identical to that of \citet{Renaud2019}, and summarized here. The simulation considered is an hydrodynamical model of an Antennae-like major merger \citep{Renaud2015}. The simulation has a resolution of $1.5 \pc$ and includes, background UV heating, atomic and molecular cooling, star formation (at an efficiency of 2\% per free-fall time above $50 \cc$, i.e. where the molecular gas fraction is expected to be close to unity at solar metallicity, \citealt{Krumholz2011}), photo-ionization, radiative pressure and type-II supernova feedbacks. 

Our simulation only comprises a limited number of feedback effects. In particular, AGN feedback is not included. Although this could affect our conclusions, in particular in the nuclear regions of the galaxies where X-ray photons could heat up the gas and excite CO \citep[e.g.][]{Krips2011}, the presence of an AGN in the Antennae has not been unambiguously established (see \citealt{Brandl2009}, and also \citealt{Ueda2012} which mention the possibility of an hidden AGN causing high line ratios in the nucleus of NGC~4039).

The results related to star formation depend on the sub-grid recipe used and details would change when using different methods. The implementation and parameters we use have been designed, at our working resolution, to reproduce observations of integrated quantities like the SFR and the depletion time of the progenitor galaxies run in isolation, typical of main sequence galaxies. The simulation also retrieves the observed star formation activity of the Antennae during the interaction (see below). Uncertainties remain on the validity of non-integrated quantities like the number, spatial distribution and masses of star-forming clouds that would change, e.g., by varying the density threshold and SFE, even if keeping the SFR unchanged. However, the morphology of the star-forming areas and the properties of the young massive cluster formed in the simulation are in line with the observations \citep{Renaud2015}, such that our models can be used to help the interpretation of observation data down to the scale of $\sim 10 \pc$.

\subsection{CO emission}
\label{sec:lvg}

The large velocity gradient (LVG) method is applied in post-processing to compute the intensity of the CO (J$\to$J-1) emission lines based on the local density and temperature of the ISM and lookup data \citep{Weiss2005}, and velocity gradients along the line of sight to infer absorption \citep[see][for details]{Bournaud2015}. The analysis is conducted at $1.5 \pc$ resolution for almost all the lines-of-sight, with rare exceptions at $3 \pc$ where the highest densities are not met. Data presented here are velocity-integrated Rayleigh-Jeans brightness temperatures, in units of $\U{K\ km\ s^{-1}}$. For convenience, we also provide the conversion to flux integral in units of $\U{Jy\ km\ s^{-1}}$. The CO-to-H$_2$ conversion factor \aco (expressed in $\msun \K^{-1} \U{km}^{-1} \U{s} \pc^{-2}$ in the rest of the paper) is inferred using the mass of the gas denser than $50 \cc$ (i.e. matching the star formation density threshold), which leads to a good agreement with empirical values for a Milky Way-like galaxy \citep{Renaud2019}. The dispersions presented are the root mean square values of a sample of 5 lines-of-sights, obtained by tilting the system by $\pm 15^\circ$ along the axes of the plane of the sky.

\subsection{Regions of interest}

\begin{figure*}
\centering
\includegraphics{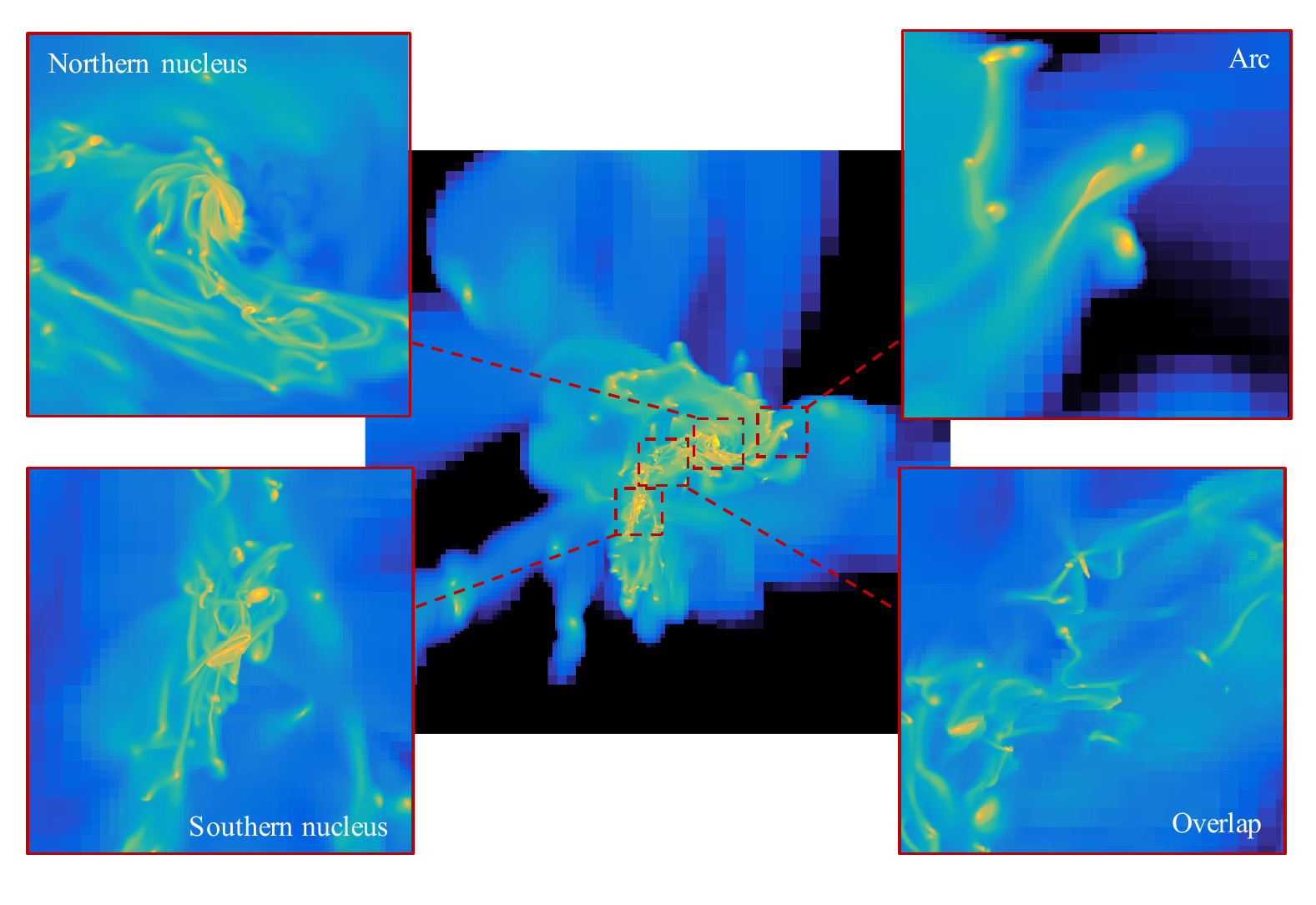}
\caption{Gas density map of the central $12 \kpc \times 12 \kpc$ of the merger. The connections with the long tidal tails are visible on the left-hand side. The smaller panels show zoom-ins in the four $1 \kpc \times 1 \kpc$ regions where our analysis is conducted.}
\label{fig:morpho}
\end{figure*}

After exploring the time evolution of the system in \citet{Renaud2014b, Renaud2019}, we now focus on the snapshot representing the best morphological match to observational data. This instant is found $5 \Myr$ after the second pericenter passage, and $19 \Myr$ before the third one which marks the onset of final coalescence. At this time, the SFR of the entire system reaches $15.7 \Msunyr$ (i.e. in the range of the observed values, $7\mh 20 \Msunyr$, see \citealt{Zhang2001,Brandl2009, Klaas2010}) and the global \aco is 2.9 (which puts the system in the post-burst regime, see \citealt{Renaud2019}).

For the sake of simplicity, we adopt a line-of-sight and an orientation comparable to that of the real system. In this snapshot, we select four star-forming regions of $1 \kpc \times 1 \kpc$, as shown in \fig{morpho}. The selection comprises the two galactic nuclei, the overlap region (on the Eastern side, left in \fig{morpho}, where the disk remnants mix) and the arc on the opposite side of the overlap in the Northern galaxy. In the real Antennae, the corresponding regions yield strong CO(1-0) emission \citep{Wilson2003}, and host or have recently hosted intense star formation activities, in particular in the form of massive star clusters \citep[$\approx 5\mh 10 \Myr$ old, see e.g.][]{Whitmore1999, Bastian2009, Herrera2012}. Our previous explorations of the system as a whole suggest that a variety of physical processes are responsible for this activity \citep{Renaud2014b, Renaud2015}. We seek here signatures of these processes in observable properties of the ISM, and in the CO emission.

For comparison, we also use another snapshot of the same simulation, during the starburst episode triggered by the coalescence, when the SFR peaks at $97.2 \Msunyr$. We refer to this point as ``ULIRG'' in the following. We also include the Milky Way-like case ($1 \msunyr$) and the gas-rich clumpy galaxy ($45 \msunyr$, representing a disk at redshift $\sim 2$) presented in \citet{Bournaud2015}.

\section{Results}

\subsection{The triggers of star formation and their delays}
\label{sec:triggers}

\begin{figure*}
\centering
\includegraphics{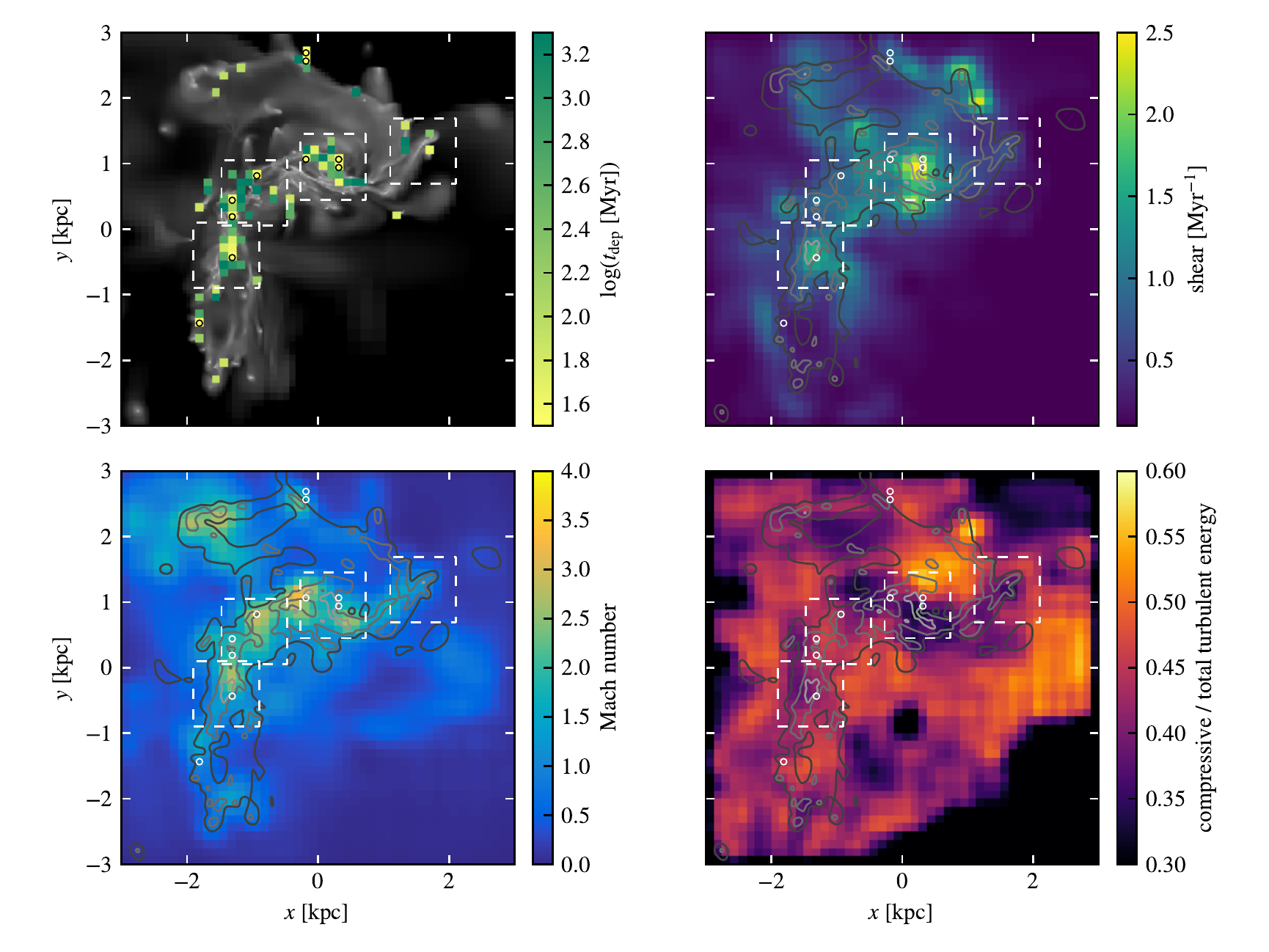}
\caption{Maps of the depletion time (top-left), shear (top-right), Mach number (bottom-left) and of the compressive to total ratio of turbulent energy (bottom-right). A gray-scale map of the gas density is underlaid in the top-left panel and is represented as contours on the others, to guide the eye. The depletion time and Mach numbers are measured at a scale of $125 \pc$, using stars younger than $10 \Myr$ for the former, while the shear and compressive to total ratio are measured at $50 \pc$ and their maps are density weighted histograms with a bin size of $125 \pc$. Small circles indicate the positions of minima of depletion times and the dashed squares show our 4 regions of interest. No clear correlations between these quantities are visible. This is due to the time-delay between the onset of a physical trigger and the resulting star formation activity being of the same order of magnitude as the dynamical time of the merger ($\sim 10 \Myr$).}
\label{fig:machmap}
\end{figure*}

We start by seeking spatial (anti-)correlations between the star formation activity and processes known to trigger (quench) it. The top-left panel of \fig{machmap} overlays the gas density map of our Antennae model with the depletion times measured at the scale of $125 \pc$, using the stars younger than $10 \Myr$.\footnote{Therefore, only the regions having hosted star formation in the last $10 \Myr$ appear in this map. A larger age cut would provide better statistics, but would also encompass stars that might have significantly drifted away from the locations (and thus the physical conditions) of their formation.}

We first find that, across the galaxies, \tdep has a scatter of several orders of magnitude, with sharp gradients especially around the overlap and the galactic nuclei (up to $\approx 20 \Myr\pc^{-1}$), as discussed in \sect{nuc}. Other regions also host a range of depletion times. The structure of the Northern arc (including our right-most region of interest and its gaseous extension toward the top-left) yields a broad distribution of depletion times, as well as dense gas clumps not hosting star formation. The map of shear (top-right panel of the figure) indicates that such clouds found at the edge of the disk remnant lie in areas of strong shear (e.g. at $x\approx 0.9\kpc$, $y \approx 2.2 \kpc$ on \fig{machmap}), likely caused by the differential tidal stretch that truncates the disks. The shear is thus destroying the over-densities and these structures will not survive long-enough to form stars. Conversely, other clouds (further away from the tidal truncation of the disk) are on the verge of hosting star formation in the next few Myr (e.g. at $x\approx -1.9\kpc$, $y \approx 2.3 \kpc$).

This indicates that the large scatter of \tdep is \emph{partly} due to gas structures being at different evolutionary stages, toward collapse or dissolution, and that the influence of the large-scale environment is blended into this ``intrinsic'' time sequence (see \citealt{Grisdale2019} on the diversity of SFEs at cloud scale). The variations of depletion times are further discussed in \sect{tdep}, in relation with the distribution of dense gas.

The four panels of \fig{machmap} illustrate the complexity of the picture by displaying the maps of some factors triggering or quenching star formation. In these maps, we retrieved expected general trends in most of the regions: volumes with strong shear have long depletion times, and those with strong compressive turbulence (i.e. high Mach number and high compressive to solenoidal ratio) tend to favor efficient star formation. Notable exceptions are the vicinity of the galactic centers, discussed in \sect{nuc}. However, a more detailed and quantitative inspection of the relations between the plotted quantities reveals no clearer trends than the general qualitative ideas mentioned above. In that sense, the causal link between these processes and star formation is not un-ambiguously established here, although it appears when considering the time evolution of the system (see \citealt{Renaud2014b}, their figure 1).

The reason for this is likely the time delay between the onset of a physical process and the measure of the resulting star formation activity. In our case, using stars younger than $10 \Myr$ implies that the delay is at least $10 \Myr$. In other words, the process that triggers (quenches) star formation, as early as the formation of a cloud, can be long gone when the depletion time is measured to be dropping (rising). This is particularly important in fast evolving systems like galaxy mergers where dynamical times are short. If causal relations are not found in our Eulerian scheme, they would likely get clearer by using Lagrangian methods (e.g. with tracers in grid codes, see \citealt{Genel2013, Cadiou2019}) and by tracking the gas material throughout the phases of compression, cooling, fragmentation and star formation \citep[][]{Semenov2017}.

Apparent in these maps is the diversity of environments leading to a broad dynamical range in the quantities plotted. For instance, the vicinity of the nuclei host strong shear and the depth of the potential implies fast moving gas structures, while the arc in the outskirts of the disk evolves with longer timescales. As a result, the relevant timescales for cloud evolution (and thus star formation) are strong functions of position and time, such that it is impossible to define global timescales. Therefore, observational tracers of any stage of the star formation process (from dense gas to gas-free star cluster) suffer from this diversity of timescales by likely introducing biases in the interpretation of galaxy-wide maps. For the same reason, the diversity of timescales translates into a diversity of scale-lengths which also introduce biases, as discussed in \sect{scale}.

According to our model, the on-going collision occurs as the two galaxies are falling back onto each other after $\sim 150 \Myr$ of separation since their first encounter. The Northern galaxy (NGC~4038) is moving southward toward the other progenitor. The overlap of the disks starts on the East side (left on \fig{morpho}) and moves roughly south-westward as the two galaxies fly through each other. The time taken by the overlap to move from one side of the stellar disks to the other is about $10 \Myr$, i.e. comparable to the timescales of the physical processes mentioned above, to the depletion times of the clouds, and to the propagation of feedback. If starbursts were exclusively triggered by shocks (when the gaseous material of the two galaxies overlap, either through cloud-cloud collisions, or between the gas reservoirs of the progenitors, see \citealt{Jog1992} and \sect{ccc}), one would expect to find a signature of this evolution with star formation running from East to West of the system. In that case, the galaxies would exhibit gradients of tracers of star formation from the signatures of the latest stages of star formation (e.g. \ha, UV, IR) toward the East, behind the shocks themselves, to more early-stage indicators (e.g. dense gas, CO, HCN, HNC, etc.) to the West. However, MUSE observations covering the entire central region reveal no such gradients in the positions or sizes of the \hii regions \citep{Weilbacher2018}, indicating that shocks are not the only triggers of the starburst activity. The observed complexity is in line with our results and we predict that similar conclusions could be reached with other tracers.

In this context, using velocity dispersion only to trace star formation might lead to errors, since an above-average dispersion could be both the cause of star formation (e.g. compressive turbulence, shocks) and its consequence (feedback). Some strong shocks not associated with young stars have been detected in the Antennae \citep{Haas2005}, but exclusively in the overlap, very likely marking the onset of localized shock-triggered star formation.

Due to this diversity in star-forming regions, their triggers and their delays, it is involved to establish clear global relations. To circumvent this, we examine now the properties of the dense ISM to identify features specific to interacting systems.

\subsection{Excess of dense gas in the star-forming regions}

\begin{figure}
\centering
\includegraphics{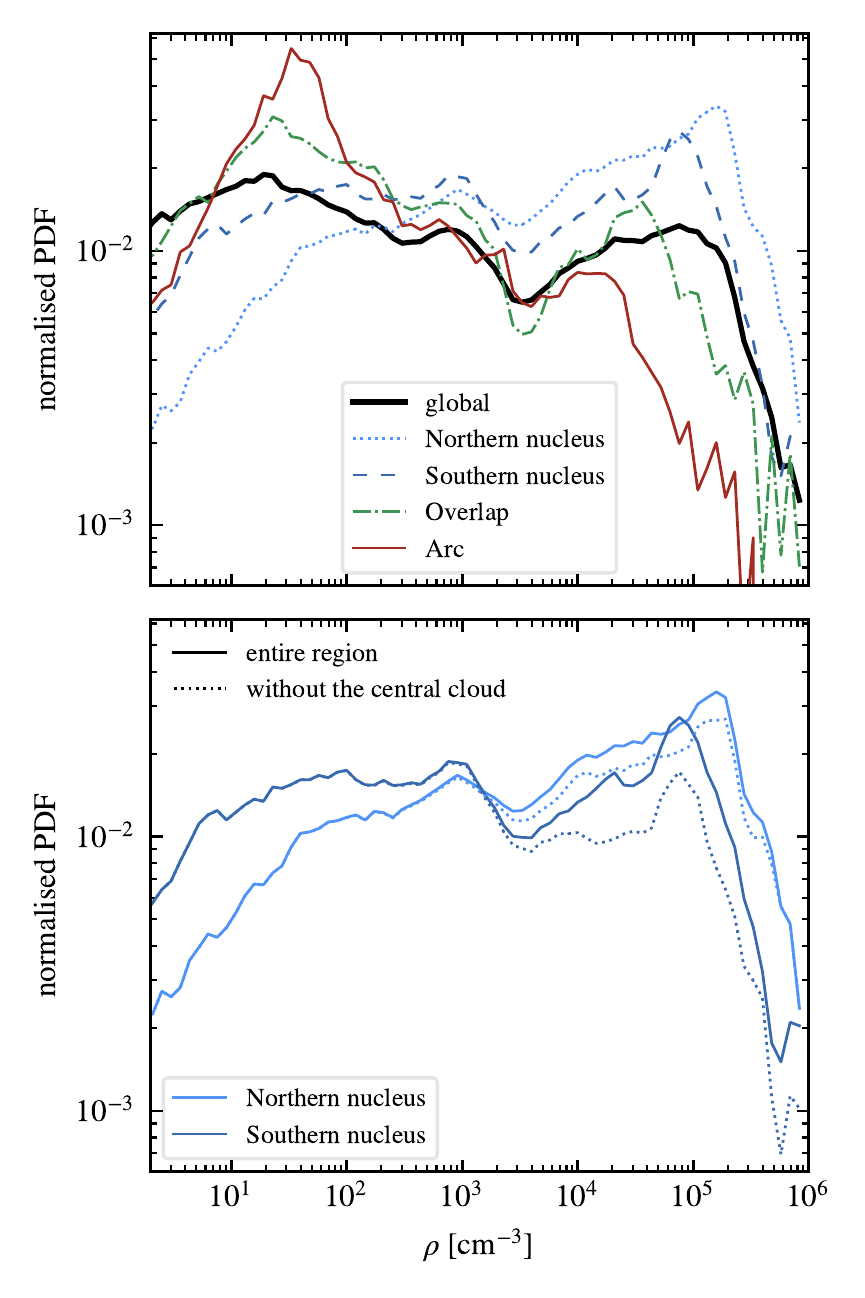}
\caption{Normalized mass-weighted gas density PDFs in the 4 regions and in the entire galaxy (top), and in the two nuclei, with and without the central clouds (i.e. the nuclei themselves, bottom). All regions exhibit a secondary peak of dense gas ($\gtrsim 10^4 \cc$), in contrast with isolated galaxies in which these highest densities only correspond to the tails of the distributions.}
\label{fig:pdf}
\end{figure}

\fig{pdf} shows the gas density probability distribution functions (PDFs) in our four regions and in the entire system. All PDFs yield a secondary peak of dense gas ($\gtrsim 10^4 \cc$), but this feature is most pronounced in the nuclei and, to a lower extent in the overlap. Together, the four kpc-scale regions gather 87\% of this dense gas across the galactic system, and only a few other clouds account for the rest (on the Eastern side of the Northern arc, and at the South-most end of the southern galaxy, recall \fig{morpho}). We note that the average density of the secondary peak varies from one region to the next. Although gas less dense than this secondary peak can also form stars, the density dependence of the star formation law ($\propto \rho^{3/2}$) makes the dense peak the main contributor ($85 \%$) to the total star formation activity.

As shown in \fig{pdf} (bottom panel), the gas clouds of the nuclei themselves (i.e. the innermost $\sim 100 \pc$) constitute a significant fraction of the mass in the secondary peak of their regions, but are not the sole contributors: the peak is still present in the PDF of the kpc-scale regions without the nuclei and its maximum is found at the same density with and without the nuclei. Star formation in the central regions is thus not restricted to the sole innermost gas structures but span numbers of other clouds with comparably high densities, as visible in the density maps (\fig{morpho}). These star-forming clouds in the immediate vicinity of the nuclei could result from the fragmentation of infalling material, but the excess of dense gas and the similarities of the PDFs with and without the nuclei suggests a more efficient process.

In addition to the presence of the peak in the overlap and the arc, this indicates that the physical origin of the secondary peak affects large-scale volumes and is not (only) related to nuclear inflows. Furthermore, \citet[their figure 2]{Renaud2014b} showed that the secondary peaks appear as early as the first passage in the interaction, and are thus not only limited to penetrating encounters, in-line with the global increase of compressive turbulence. The link between the shape of these PDF and turbulence is further discussed in \sect{forcing}.

It is important to note that these PDFs do not result from a shift nor a widening of those found in isolated galaxies: the presence of a secondary peak is unique to interactions and reflects that star formation in such systems is not a simple extrapolation of that of isolated galaxies. Therefore, the physical reason for the enhanced star formation, both in term of rate and efficiency (i.e. SFR and \tdep), lies in the process(es) that generate this excess of dense gas. In the next section, we connect this excess to reduced depletion times.

\subsection{Variations of depletion time}
\label{sec:tdep}

At the instant considered, the entire galactic system has a depletion time of $100 \Myr$ ($180 \Myr$, if considering the molecular and atomic gas), i.e. about one order of magnitude shorter than that of normal star-forming disk galaxies, which places it in the regime of LIRGs/ULIRGs defined by \citet{Daddi2010b}.

Our kpc-sized regions have depletion times of $65 \Myr$ (resp. 120) in the overlap, $80 \Myr$ (resp. 120) in the nuclei and $170 \Myr$ (resp. 250) in the arc (when accounting for the molecular phase only, and the total gas mass respectively). These measurements are compatible with the estimates of \citet{Bigiel2015} who reported shorter depletion times when using tracers of dense gas ($\sim 40\mh 160 \Myr $ in total infrared) and longer times with CO, but at slightly larger scales ($\approx 2\mh 3 \kpc$) and by adopting disk-like luminosity-to-mass conversion factors (while the \aco model from \citealt{Renaud2018} would lead to depletion times about 1.5 shorter). Already by selecting the regions hosting the most intense star formation activity, we report significant variations of \tdep, but these 4 regions are not representative of all the star-forming areas, and thus we also explore the variation of \tdep in a more systematic way, and with a larger statistical sample, on a cloud-basis.

\begin{figure}
\centering
\includegraphics{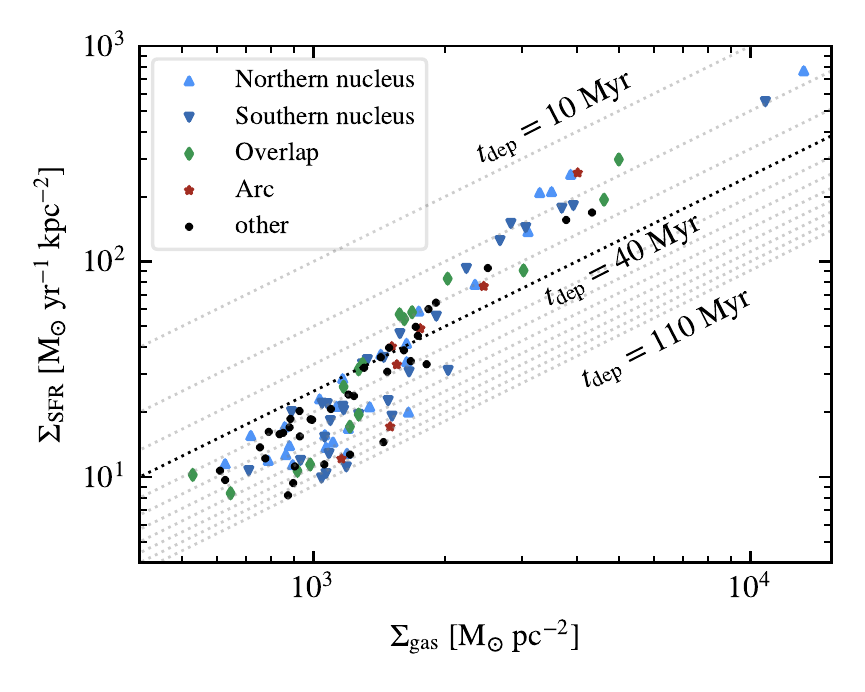}
\includegraphics{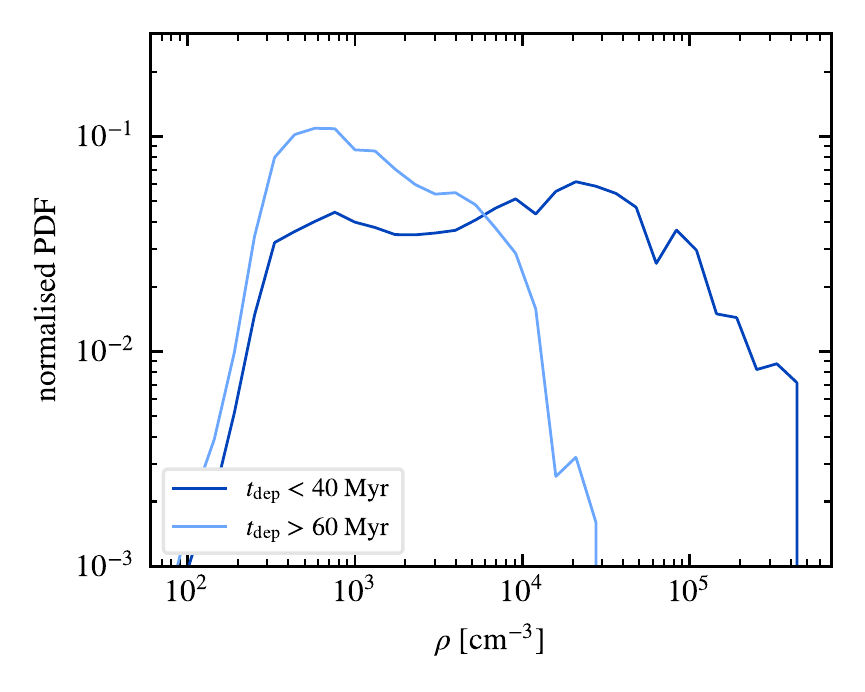}
\caption{Top: Schmidt-Kennicutt diagram of the clouds identified in the simulation. The two densest points correspond to the clouds at the galactic centers. The dotted lines mark constant depletion times (from 10 to $110 \Myr$, in steps of $10 \Myr$), with the dark one corresponding to $40 \Myr$. Bottom: Normalized mass-weighted gas density PDF of these clouds, split into two regimes of depletion times: < 40 and > 60 Myr. Clouds with 40 Myr < \tdep < 60 Myr correspond to the transition regime shown in the top panel and are not considered here for the sake of clarity. The distribution of clouds does not follow a unique power-law, but rather a combination of two regimes: clouds with an excess of dense gas have a shorter depletion time than those without.}
\label{fig:ks}
\end{figure}

The top panel of \fig{ks} shows the positions of star-forming clouds in the Schmidt-Kennicutt diagram. The surface densities are computed at the scale of each cloud ($\sim 30 \pc$), independently of each others. The overall relation is super-linear as the points are best-fitted by power-laws of indexes 1.4 when including the 2 centermost clouds, and 2.0 when excluding them, but with large deviations (69.2 and 19.2 respectively, in the units of the figure).

Instead of interpreting this distribution as a unique power-law with a large scatter, we highlight that a line of constant depletion time splits the clouds into two families: those with short depletion times ($\tdep \lesssim 40 \Myr$) are found in the dense-end of the distribution while the others are systematically less dense. The dichotomy directly relates to the PDFs of these clouds, shown in the bottom panel of \fig{ks}. Clouds with short \tdep systematically yield an excess of dense gas ($\gtrsim 10^4 \cc$) with respect to the others. More than the mere expected relation between short \tdep and high densities, this shows that all the clouds with $\tdep < 40 \Myr$ contain an excess of dense gas, while none of the others do. Furthermore, not all star-forming clouds but only those with a short \tdep contribute to the secondary peak noted in \fig{pdf}, demonstrating that the excess of dense gas noted before as the presence of a secondary peak is associated with short depletion times, in addition to high SFRs.

At small scale, the density of star formation rate depends on the gas density in a super-linear fashion (e.g. $\rho_{\rm SFR} \propto \rho^{1.5}$ in the case of constant efficiency per free-fall time, as adopted here). This implies that, for a given gas mass, an excess of dense gas (our secondary peak) leads to a more efficient star formation (i.e. a decrease in depletion time).

Both the Schmidt-Kennicutt diagram and the distribution of \tdep show that the interacting galaxies can contain \emph{simultaneously} clouds with unimodal PDFs and long depletion times (in line with those of isolated galaxies), and clouds with an excess of dense gas and significantly shorter depletion times. The mix of both types sets the properties of the galaxies at kpc-scale which, in our case at this instant, puts the Antennae among the galaxies with shortened depletion time, in the starburst regime.

Comparable variations of \tdep have also been found observationally across the Antennae \citep{Bigiel2015}, IC~4687 \citep{Pereira2016} and NGC~2276 \citep{Tomicic2018}. In the Antennae (the observation and our model), the complexity of the geometry due to the deformation from the previous interaction and the partial overlap of two galaxies lead to complex galactic-scale gradients, neither radial, nor from one side of the galaxy to the other (unlike in NGC~2276 where ram pressure and tides imprint clear variations of \tdep, \citealt{Tomicic2018}). The absence of such large-scale, smooth variations indicates that the triggers of star formation are a complex interplay of physical processes, and/or of processes with timescales or the order of the dynamical times of the clouds, as hinted in \fig{machmap}.

The exact physical origin of this dense gas is difficult to assess in our models, as discussed in \sect{triggers}. However, some additional clues can be obtained by examining the star formation histories of the clouds. Clouds with long depletion times ($\tdep > 40 \Myr$) are forming less stars now than $5 \Myr$ ago. This indicates that their formation activity is slowing down, and will probably stop within a few Myr, due to the consumption of dense gas and/or its dispersion by feedback. The picture is less clear for the clouds with shorter \tdep. Some show an increasing star formation activity while others behave like the longer \tdep objects mentioned previously. This points toward a mix between evolutionary and environmental effects shaping the star formation activity of these clouds (see \citealt{Grisdale2019}, in the context of the Milky Way.). 

Interestingly, the locii of the clouds in the Schmidt-Kennicutt diagram (\fig{ks}) is independent of their host region, and more generally of their position in the galaxies. In particular, we find dense clouds with short depletion times in all the regions, which suggests that differences in the kpc-scale environment play a minor role on the properties of the clouds themselves. The other possible explanation is that the diversity of processes leading to short depletion times across the galaxies are indistinguishable in a Schmidt-Kennicutt diagram (with the exceptions of the two nuclei which reach extreme densities, but comparable depletion times as many other clouds). 

\subsection{Scale dependence}
\label{sec:scale}

\begin{figure}
\centering
\includegraphics{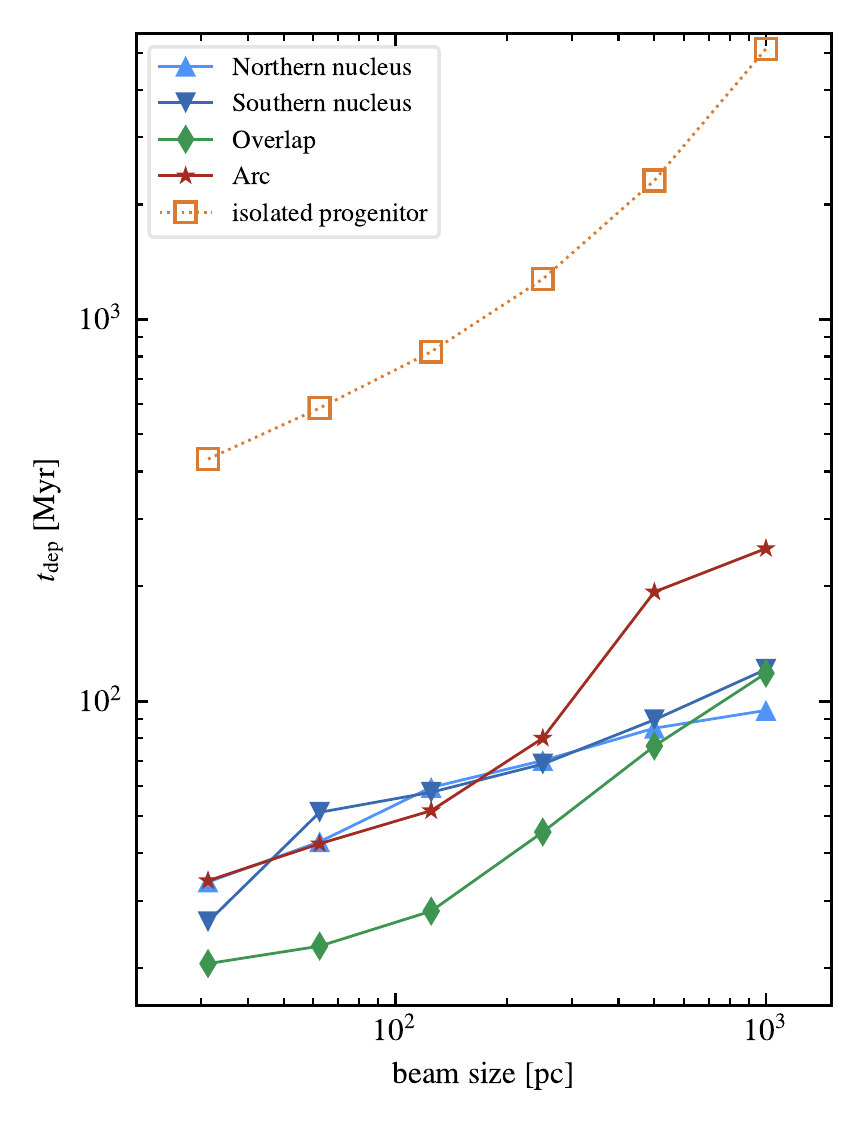}
\caption{Depletion times as a function of the scale over which they are measured, and that of one of the progenitor galaxy, run in isolation for reference. While the general trend is common to all regions, details depend on the sub-kpc structure of the ISM and significantly vary from one region to the next, due to different organizations of the clouds set by the kpc-scale environment.}
\label{fig:tdepbeam}
\end{figure}

In \fig{tdepbeam}, we consider different areas (or ``beam'' size), centered on the star formation peaks within our four regions of interest. Due to the spatially extended enhanced star formation activity, any of our regions yields a depletion time an order of magnitude shorter than that of equivalent volumes in a normal star-forming disk ($\sim 1 \Gyr$, as expected). Not surprisingly, increasing the beam size leads to, on average, including a larger fraction of non-star-forming gas, which increases the depletion time \citep[see e.g.][]{Schruba2010}. Details of this relation however depend on the size of individual star-forming structures and the separation between them within the beams, and the latter varies across the galaxies due to the diversity of processes involved (recall \fig{machmap}). This is illustrated by the differences between our four regions of interest in \fig{tdepbeam}, on top of the expected general trend of increase of \tdep with the beam size, as also seen in the isolated galaxy.

For instance, star formation in the arc occurs in a handful of clouds, surrounded by non-star-forming, diffuse material. Therefore, the depletion time in this area rapidly increases when it is measured at scales larger than the size of these clouds. In the other regions, this effect is still visible but without a particular scale identifiable due to the larger filling factor of clouds in the area\footnote{The mean separation between clouds is $270 \pc$ in the arc and $120\mh 140 \pc$ in the other regions. The clouds have different shapes (for instance, some are elongated due to shear), but their volume is approximately constant across the galaxy, corresponding to spheres of radius $30\mh 35 \pc$.}, which means that increasing the beam size leads to including more clouds, i.e. a lower fraction of non-star-forming material than in the arc. 

\begin{figure}
\centering
\includegraphics{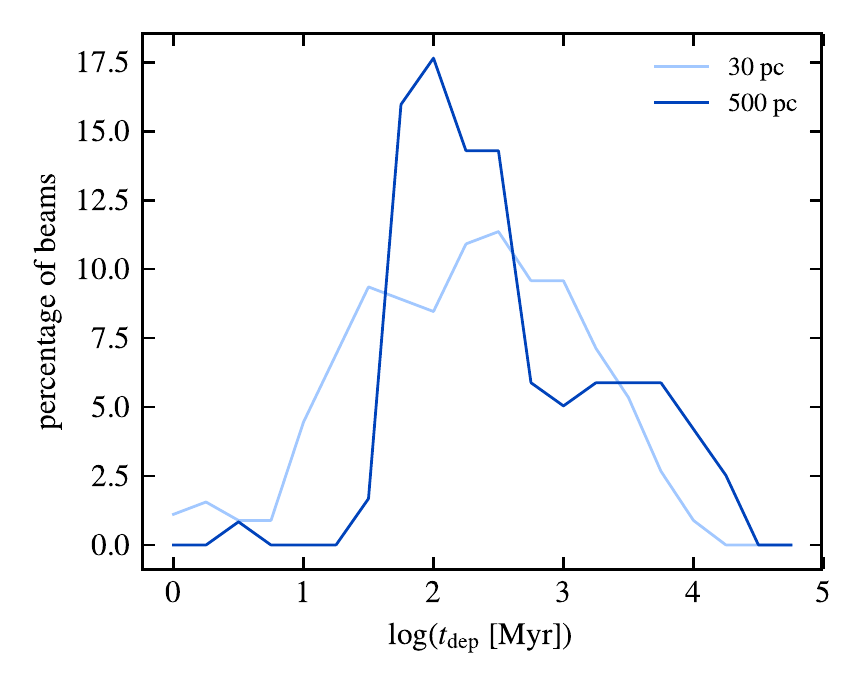}
\caption{Distribution of depletion times, measured in 30 pc and 500 pc beams on a Cartesian grid (with arbitrary origin, i.e. not centered on peaks of star formation) in the central $6 \kpc \times 6 \kpc$ of the system. For better statistics, we have repeated the measurements by shifting the origin of the grid by several fractions of the beam size, and stacked the distributions obtained. Large scale measurements have a bimodal distribution of short ($\sim 100 \Myr$) and long ($\sim 2 \Gyr$) \tdep.}
\label{fig:tdepdistrib}
\end{figure}

\fig{tdepdistrib} shows the distribution of depletion times when measured at cloud (30 pc) and galactic (500 pc) scales. At large scale, the regions of shortest depletion times are blended with less and no star-forming material, which truncates the distribution on the short \tdep side. For the same reason, regions not forming stars ($\tdep = \infty$) are binned with regions with long (but finite) depletion times ($\gtrsim 5 \Gyr$), which extends the distribution to its long side with respect to that at small scale. As expected, using beam sizes larger than star-forming regions dilute them towards longer \tdep.

The galactic-scale distribution reveals the two modes of depletion times: the main one at $\sim 100 \Myr$ and a secondary at $\sim 2 \Gyr$. These average \tdep's approximately correspond to that of the disk and starburst regimes found in Schmidt-Kennicutt diagrams \citep[e.g.][]{Daddi2010b}. At cloud scale, however, the distribution is narrower (although it appears wider due to the log-scale of \fig{tdepdistrib}) because of the lack of the dilution effect mentioned above. However, it still spans several orders of magnitude in depletion times, despite the uniform star formation recipe at constant SFE imposed (0.02 at $1.5 \pc$). Details on the physical origin of this scatter are discussed in \citet{Grisdale2019} in the context of more quiescent disk galaxies, and Kraljic et al. (\inprep) in the case of interacting systems. 

In summary, all regions contain clouds with short and long depletion times, with an excess of dense gas or not. This implies that the galactic-scale dynamics has either no influence on the depletion times of individual clouds and thus that the scatter is a purely intrinsic evolution, or that the different dynamical triggers at stake have the same signatures on the depletion times. However, the details in the scale dependence of \tdep reveal a different organization of the star-forming material from one region to the next (in particular in the arc), linked with the kpc-scale triggering of cloud formation.

\subsection{Long depletion times in the nuclei}
\label{sec:nuc}

\fig{tdepbeam} shows that the depletion time yields a steeper dependence with scale for beam sizes $\lesssim 100 \pc$ in the nuclei than in the overlap and the arc. The reason for this originates from the peculiarity of the nuclei in the gravitational potential of the galaxies. Being natural convergence points of flows and minima in the potential, they induce strong shearing and tidal effects along the fueling channels toward them (recall \fig{machmap}). These effects smooth out most (but not all) of the gas clumps, and in turn prevent their collapse on over-dense seeds. Consequently and despite large amounts of dense gas, star formation is considerably slowed down a few $\sim 100 \pc$ away from the centers. However, when the gas reaches the nuclei themselves ($\sim 50\mh 100 \pc$ in our simulation), it accumulates, leading to important over-densities and the observed star formation activity. As a result, efficient star formation is concentrated in the main structure of the nuclei, while it is more evenly distributed in the overlap, due to the large shocked and compressed volumes. This situation is qualitatively comparable to that of barred galaxies where the bar(s) generate the torques necessary to fuel large amounts of gas, but where shear increases \tdep of this gas, or even quenches star formation \citep[see][in the context of the Milky Way]{Emsellem2015, Renaud2015d}, with the notable difference that the excess of dense gas in the Antennae allows some clouds to survive destruction due to their higher binding energy (recall \fig[s]{morpho} and \ref{fig:machmap} and see \sect{ccc}).

This point explains why observations focusing on the nuclei can report longer depletion times there than in the overlap, despite higher SFRs (see e.g. \citealt{Bigiel2015}). Our results predict that observing these regions with beams comparable to the sizes of the star-forming area only (i.e. encapsulating as little surrounding non-star-forming material as possible) would lead to the expected shorter \tdep in the nuclei. We also note that the scale dependence of the depletion time varies with the clumpy nature of star formation, which itself depends on the tracer used to measure the SFR. For instance, a tracer of molecular gas is expected to reflect larger objects and thus milder variation of \tdep with the scale than, e.g., the \ha luminosity which is more concentrated around the star-forming sites.

\subsection{CO emission}

\subsubsection{Spectral line energy distributions}
\label{sec:sled}

\begin{figure*}
\centering
\includegraphics{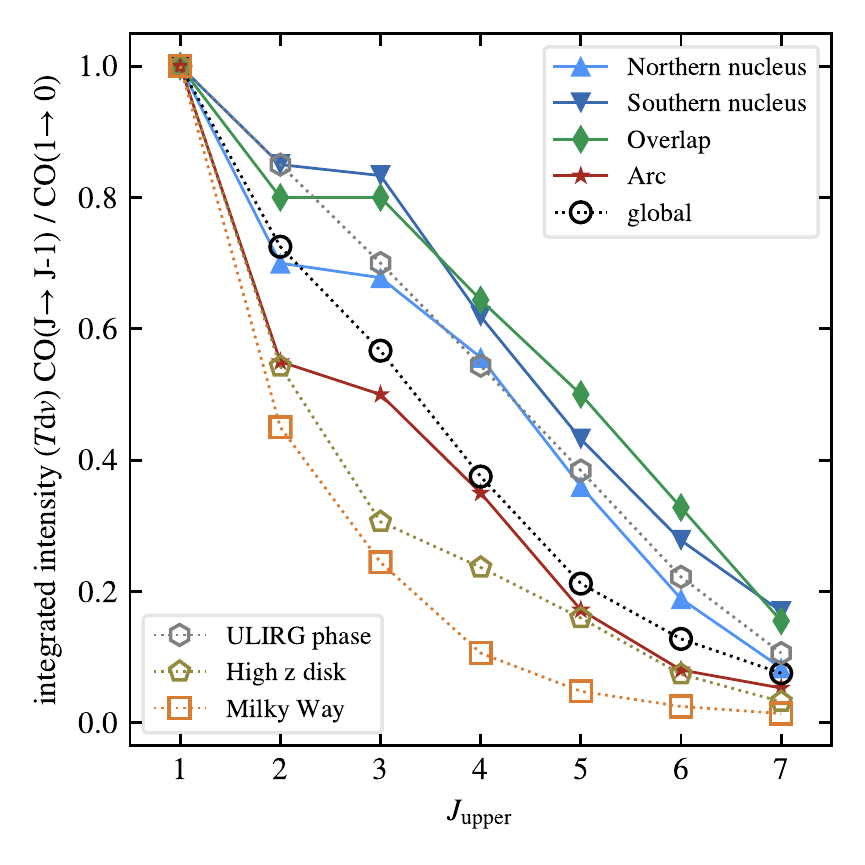}
\includegraphics{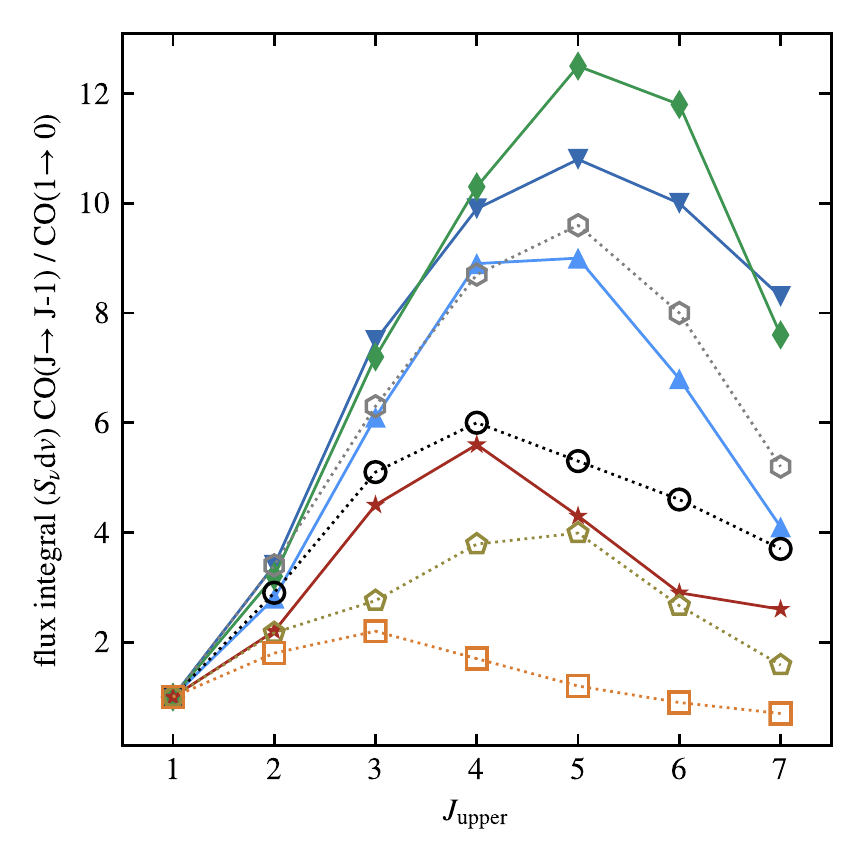}
\caption{CO SLEDs for velocity-integrated brightness temperature (left) and velocity-integrated flux density (right), both normalized to CO(1-0), in our 4 selected regions (filled symbols), in the entire system (``global''), in the same simulation but during the starburst episode (ULIRG), in a gas-rich clumpy disk (high-redshift) and in a Milky Way-like galaxy (open symbols). The SLEDs of the most actively star-forming regions (nuclei and overlap) resemble that of the Antennae (global and ULIRG), while the arc better matches the high redshift disk at high \jup.}
\label{fig:sled}
\end{figure*}

\fig{sled} shows the spectral line energy distributions (SLEDs), in term of integrated intensity and flux integral, of our regions ($1 \kpc$) compared to entire galaxies ($10\mh 30 \kpc$): the Antennae at the same time (named ``global''), at the time of maximum starburst (named ``ULIRG''), a Milky Way-like galaxy, and a high redshift clumpy disk. 

Our regions of interest encompass a range of densities, from the peaks in the clouds, to the more diffuse warm medium in between them. For this reason, we expect our modeled CO emission to differ from that observed when exclusively focusing on clouds. Our results would thus better match observations with beams not resolving the individual clouds and thus blending the peaks of cold emission with warmer media, i.e. on scales of the order of and larger than $\sim 1 \kpc$. Indeed, while our results for the Antennae are in qualitative agreement with the observations of \citet[$\sim 0.6 \kpc$ beams, centered on peaks]{Bigiel2015}, they better match those of \citet[$\sim 4 \kpc$ beams]{Schirm2014}, as discussed in \app{obs}. All our 4 regions and the Antennae system (global and ULIRG in \fig{sled}) yield a strong CO(3-2) emission, scattered around CO(3-2)/CO(1-0) $\approx 0.6$, which is similar to expected values in (U)LIRGs \citep{Yao2003, Papadopoulos2012, Bolatto2015}.

The high CO(3-2) intensity (in particular the ratio of brightness temperatures CO(3-2)/CO(2-1)) in the kpc-scale regions is not retrieved in any of the large scale measurements, regardless of the star formation activity. A similar trend might be seen in the observational SLEDs from \citet{Bigiel2015} who used a beam size smaller than our regions ($\approx 0.6 \kpc$). This suggests that such emission is localized in small volume(s) and thus gets blended with regions of weaker emission at larger scales. This may correspond to the CO(3-2) detection in the real system of a bright super giant molecular cloud (the so-called ``firecracker'') supposedly being the future formation site of a massive star cluster (\citealt{Whitmore2014}, see also \citealt{Johnson2015}). 

At higher excitation levels ($\jup \ge 4 \mh 5$), the nuclei and the overlap regions have SLEDs comparable to that of the ULIRG phase, but the arc (and the galactic-scale ``global'' measurement) resembles more the high-redshift clumpy galaxy, with a significant bend in their SLEDs (in integral intensity). We propose an interpretation of these features in \sect{ismstructure}.

\begin{figure}
\centering
\includegraphics{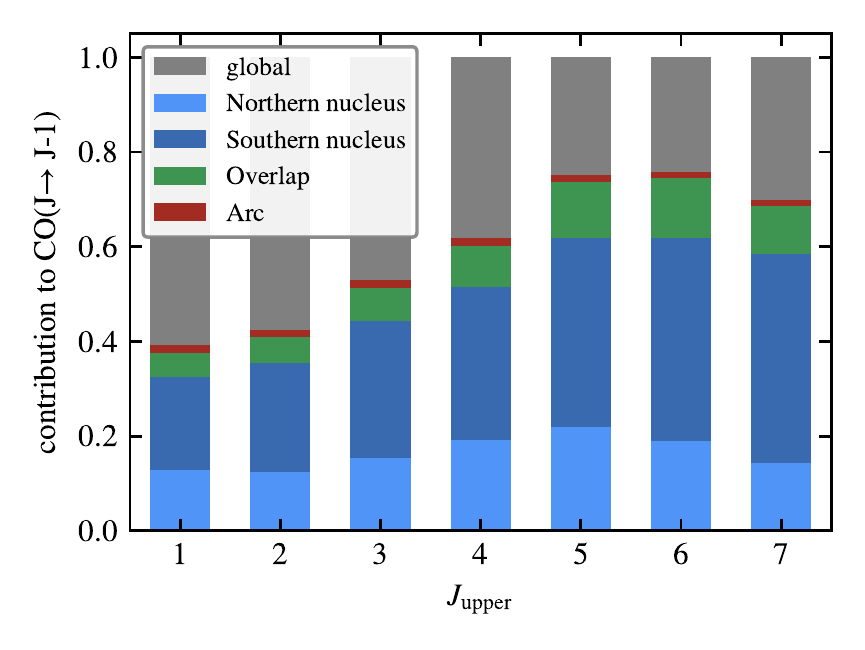}
\caption{Contribution to the CO(J$\to$J-1) line from our four $1\kpc \times 1\kpc$ regions, with respect to the entire system (``global'', i.e. $10 \kpc \times 10 \kpc$). The relative contribution of our four regions varies with \jup due to differences in the density, temperature and velocity dispersion of the ISM, particularly in the inter-cloud medium.}
\label{fig:cocontrib}
\end{figure}

\fig{cocontrib} shows the contribution of the regions (1 kpc scale) to the global CO(J$\to$J-1) intensity (10 kpc scale). As expected, each region contributes more than average to the global intensity, for all transitions. Their combined contribution ranges from $\approx 40\%$ (CO(1-0)) to $75\%$ (CO(6-5)) of the total intensity, and 85\% of the SFR, although they only span $4\%$ of the area considered. The remaining, yet significant, fraction of the flux originates from less concentrated regions.

The relative contribution of the individual regions increases with \jup up to 6 for the nuclei and the overlap. However, that of the arc peaks at CO(4-3), which further tells apart the arc. As shown in \fig[s]{sled} and \ref{fig:cocontrib}, the Southern nucleus is more excited and has a higher CO intensity than the northern one at all \jup, despite hosting lower amounts of dense gas and with lower maximum densities (\fig{pdf}), and a comparable distribution of temperatures. One would then expect a stronger emission from the Northern nucleus, contrarily to our measurements. However, it hosts a stronger shear and higher Mach number (\fig{machmap}) than its southern counterpart, i.e. an overall larger gas velocity dispersion, which accounts for a stronger absorption of the CO flux captured by the LVG method. The exact relative roles of the gas density, the temperature and the velocity dispersion are difficult to assess due to sharp variations of these quantities on small-scales and the limited resolution of our analysis. Our results on the differences between the regions presented in the previous sections highlight that the relative importance of each quantity varies across the galaxy (and thus likely in time too), in a complex manner (see also \sect{ismstructure}). 

\subsubsection{Spatial variations of \aco}
\label{sec:aco}

\begin{figure}
\centering
\includegraphics{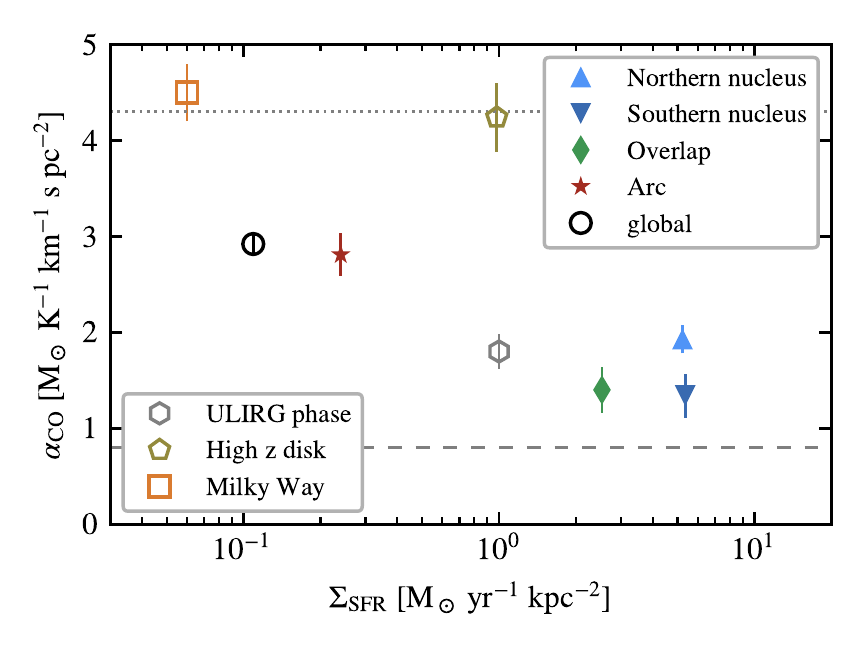}
\caption{$\alpha_\textrm{CO}$ as function of the surface density of SFR, in the four selected regions ($1 \kpc \times 1 \kpc$, filled symbols), for the entire merger system (``global''), the same system during the ULIRG phase, a gas-rich clumpy disk (high-redshift) and a Milky Way-like galaxy (open symbols). Error bars show the dispersions of the values, evaluated by varying the line-of-sight (recall \sect{lvg}). Horizontal lines mark the values commonly used in observations (4.3 for disks, \citealt{Bolatto2013}, and 0.8 for ULIRGs, \citealt{downes1998}). Important variations of \aco are found across the galaxies, due to different interplays of physical processes setting the assembly of star-forming regions, and making our regions lie in between the values traditionally assumed for disks and ULIRGs.}
\label{fig:aco}
\end{figure}

\fig{aco} shows the CO-to-molecular conversion factor \aco in our regions. As noted in \citet{Renaud2019}, the SFR and its surface density are not good tracers of variations in \aco, and the recent SFH must also be considered, in particular to account for the effects of feedback on temperature and velocity dispersion. When doing so, we find that all regions contain stars formed during the first passage ($\approx 150 \myr$) and a recent boost of the activity ($\approx 20 \Myr$) due to the second encounter. Tracking backward in time the elements constituting our four selected regions reveals that the arc is made exclusively of material from the outskirts of the disk, while the overlap comprises material from a wider range of radii, including only a couple of kpc from the galactic centers. The difference in gas densities between these volumes explains that, during the separation of the galaxies, star formation slows down significantly more in the future arc material than in the elements ending up in the other regions. This translates into a deficit of $20\mh 100 \Myr$ old stars in the arc compared to the other regions. Furthermore, tides, shear, shocks, and the resulting turbulence reshape the structures of clouds, particularly in the nuclei and the overlap, which shifts young stars away from the densest gas. As a result, stellar feedback can propagate further and with less resistance to large scales in the nuclei and the overlap (see \sect{coldism}). Therefore, the ISM of the nuclei and the overlap contain more gas of intermediate densities and warm temperatures ($\sim 10 \cc$, $10^4 \K$) than the arc. The media in which CO emission propagates are thus different in the arc and the other regions, which translates into a different absorption and in turn, into a different \aco. The same argument explains the slight difference between the two nuclei (recall also \sect{sled}).

The new onset of star formation triggered by the second passage ($\sim 5 \Myr$ ago) is too recent to have significantly altered the CO emission and propagation over kpc-scales (see \citealt{Renaud2019}) and therefore only the older activity affects our results. The galaxies thus lie on the post-burst regime and with slightly higher \aco's in the outskirts of the disks than in the central regions.

As noted in \citet{Renaud2019}, \aco for the entire system runs from 4.27 in the isolated phase to 1.78 at the peak of the starburst triggered by the final coalescence, i.e. a variation by a factor of 2.4 along the course of the merger. This is comparable to the variation between our regions of interest from our selected snapshot (from 1.3 to 2.8, i.e. a factor 2.2, see \fig{aco}). However, we note that we have focused our analysis on actively star-forming regions, and that \aco is likely much higher in other areas. Therefore, we expect variations of \aco across space to be possibly larger than those along time, even when a galaxy goes from the disk regime to the starburst one, which could have important implications on the interpretation of observations resolving the diversity of star-forming environments, even across a given galaxy.

\section{Discussion}

\subsection{Compressive turbulence and the shape of the density PDF}
\label{sec:forcing}

Sub-kpc scale simulations of the ISM are commonly used to study the inner physics of molecular clouds, down to the individual star level \citep[see a review in][]{Hennebelle2012}. These simulations do not include the galactic scales from which part of the turbulence is generated ($\gtrsim 0.1 \mh 1 \kpc$, see \citealt{agertz2009, Bournaud2010b, Renaud2013b, Falceta2015, Grisdale2017}), and they must resort to manually forced turbulence. Such works indicate that the turbulent ISM yields a log-normal density PDF \citep[see e.g.][]{Vazquez1994} of which the variance reads:
\begin{equation}
\sigma^2 = \ln{\left(1+b^2\mach^2\right)},
\label{eqn:forcing}
\end{equation}
with $\mach$ the Mach number, and $b$ a dimensionless parameter related to the forcing nature of the turbulence, i.e. the mix between its compressive (i.e. curl-free) and solenoidal (i.e. divergence-free) modes \citep[see e.g.][]{Kritsuk2007, Federrath2010}. In these simulations, the PDF is found to always have a log-normal shape\footnote{A log-normal density PDF may arise artificially from observations due to incompleteness of the lowest column density bins truncated by a limited field of view \citep{Alves2017}. The log-normal shape is however retrieved when considering large enough areas to encompass the low density envelops of clouds in closed contours \citep{Kortgen2019}.}, regardless of the turbulence forcing (with a power-law tail at high densities when self-gravity is captured, \citealt{Ballesteros2011, Elmegreen2011, Renaud2013b}). In other words, for a given Mach number these simulations report a wider, but still log-normal, PDF in compressive turbulence than for equipartition. 

As presented in \citet{Renaud2014b} and \fig{pdf}, the shape of the density PDF in our interacting galaxy models significantly deviates from the classical log-normal shape during the interactions, with the formation of a secondary peak at high densities \citep[see also][]{Teyssier2010, Bournaud2011b}. The same bimodal feature is also seen in real interacting galaxies, in term of line-width distributions \citep{Sun2018}, further suggesting the superposition of a high $\sigma$ mode to the classical one in efficiently star-forming systems. \citet{Renaud2014b} attributed this excess of dense gas to turbulent compression that arises during the interaction: changing the turbulence from energy equipartition between the compressive and solenoidal modes before the interaction to a compressive-dominated turbulence until the post-merger phase increases the gas density contrast and generates this excess of dense gas. Therefore, a discrepancy exists between the log-normal PDF from forced compressive turbulence in small-scale simulations, and the double-peaked one measured in the compressive turbulence of galaxy mergers (as is, i.e. without any manual forcing).

In the context of the Schmidt-Kennicutt relations, the analytical model of \citet{Renaud2012} uses the secondary peak to explain the observed differences in depletion times between disks and starbursting galaxies. In this model (and others of the same nature, see e.g. \citealt{Padoan2011, Hennebelle2011, Federrath2012}), using a log-normal PDF shape for all galaxies would lead to a unimodal star formation law, in which the SFR is set by the width of the PDF (\eqn{forcing}), and would thus not retrieve the difference between a gas-rich isolated disk with a high SFR (e.g. BzK galaxies, \citealt{Daddi2010}) and a local (U)LIRG with the same SFR \citep{Kennicutt1998}. In other words, increasing the variance of the log-normal PDF (either by increasing $\mach$ or $b$ in \eqn{forcing}) would only shift a galaxy along the canonical Schmidt-Kennicutt relation, but keeping it on the relation. Such a model could retrieve a diversity of SFRs, but not the diversity of depletion times. Another effect is needed to change the regime, e.g. moving from the relation of disks to that of starbursts by decreasing the depletion time. 

As discussed in \sect{tdep}, the super-linearity of the star formation law linking the local density of SFR to that of gas is the key to change the regime, i.e. to move the galaxy from the relation of disks to that of starbursts. With a wide, unimodal PDF, the densest gas constitutes the tail of the PDF, and increasing its width lead to an higher SFR. Conversely, the presence of a secondary peak in the distribution (i.e. the excess of dense gas we report) and the super-linearity of the star formation law imply that most of the star formation activity takes place at the highest densities, therefore with a higher efficiency than usual (i.e. a shorten depletion time). (See \citealt{Renaud2012} for an analytical demonstration, and recall \fig{ks}.)

We thus report a fundamentally different shape of the PDF between that from manual forcing of compressive turbulence in sub-kpc scale simulations, and that of compressive turbulence consistently arising in our simulations of interacting systems. We do not fully understand the reason of this difference but, assuming it originates from the different setups of the simulations, we postulate it could be related to the injection scale of the turbulence, its forcing mechanism and/or its dissipation. It could also be that galactic simulations contain large gas reservoirs to replenish the PDF when compression occurs, while simulations at cloud scales would suffer from a rarefaction of gas to populate both our two peaks. By nature, galactic-scales simulations like ours are likely better capturing the injection and forcing mechanisms, but the limited resolution implies the dissipation is less accurate than in small-scale studies. Further investigations, with a large dynamical range are required to conclude.

The exact nature of turbulence has important implications on the structure of the ISM and the propagation of feedback, potentially down to the stellar initial mass function (IMF) itself. For instance, theoretical models predict that a compressive turbulence would lead to a bottom-heavy IMF \citep{Chabrier2014}. However, \citet{Renaud2014b} have established the existence of compression-dominated turbulence in interactions only at the scale of $\sim 40 \pc$, i.e. several orders of magnitude larger than that of pre-stellar cores. Therefore, there are still many uncertainties on how this type of turbulence, driven by galactic dynamics, cascades and potentially dissipates down to the core scale, in addition to our lack of understanding of the core-to-star mapping \citep{Goodwin2008, Motte2018}. Small-scale studies with detailed implementations of a range of physical processes conclude on the independence of the IMF with environmental conditions, in particular the turbulence and the metallicity \citep{Bate2009, Bate2019, Myers2011}. However, including self-consistently the full range of galactic-driven effects, in particular turbulence injected at $\gtrsim 100 \pc$ scale, seems essential to capture and predict the possible variations of the IMF. This remains out-of-reach for present-day models and supercomputing resources.

\subsection{The galactic nuclei as collision environments}
\label{sec:ccc}

All the aspects of our analysis reveal similarities between the overlap and the nuclei, suggesting that the former could be a scaled-down version of the latter, i.e. with similar physical processes active, but at lower rates. From a morphological point of view, the two types of regions show several dense, cold star-forming clumps, with an average separation of $\approx 130 \pc$, distributed along elongated structures of warm gas at intermediate densities ($\sim 10^{2\mh 3}\cc$, $\sim 10^{1\mh 2} \K$), themselves surrounded by more diffuse medium ($\sim 1\mh 10 \cc$, $\sim 10^{2\mh 4} \K$).

The overlap is, by definition, the location of kpc-scale shocks between the gas reservoirs and the clouds (and/or marginally stable over-densities) they contain \citep{Jog1992}. As simulated and observed in other contexts, cloud-cloud collisions trigger star formation \citep{Tan2000, Tasker2009, Inoue2013}, in the form of massive clusters (due to the enhanced external pressure \citealt{Elmegreen1997, Fukui2014}) and possibly massive stars \citep{Motte2014, Takahira2014}. Although it is difficult in our Eulerian simulation to track individual gas structures and establish collision rates, the presence of thin and dense tidal tails and bridges connecting the density peaks in the overlap confirms the ubiquity of cloud-cloud collisions (\fig{morpho}). It is thus likely that a significant fraction of the star formation activity in the overlap is linked to cloud-cloud or cloud-reservoir collisions.

In the nuclei, the gas is fueled toward the center by negative gravitational torques, mainly from one galaxy on the other \citep{Keel1985, Hernquist1989, Barnes1991}, but also from asymmetric structures \citep[see e.g.][]{Emsellem2015}. The clumpy nature of the fueled material (recall \sect{nuc}) makes the nuclear accretion comparable to a series of cloud-cloud collisions. In that sense, the nuclei share a common mechanism with clouds in the overlap region, but likely at a higher rate due to their preferential location in the potential.

In both the overlap and the nuclei, the trigger(s) of star formation is a short-range effect, of which intensity peaks during penetrating galaxy encounters (but remains relatively high during separation phases due to the high gas density in these inner galactic regions), which then explains that the \aco's of these areas are closer to the starburst regime highlighted in \citet{Renaud2019}, than the transitional post-burst regime as it is clearer for the arc.

In short, the nuclear inflows induced by gravitational torques during close encounters play a similar role as the overlap of the outer disks, by triggering collisions between dense gas structures that lead to a starburst activity lasting $\sim 100 \Myr$ \citep{Renaud2014b}. The similar physical processes translate into SLEDs and \aco factors in the galactic nuclei and the overlap resembling those of starbursting systems, while the activity in the arc appears to have fundamentally different signatures, closer to that of disks with a high but steady SFR.

\subsection{The imprint of the ISM structure on the CO emission}
\label{sec:ismstructure}

By fitting separately the SLEDs from the cold and warm phases, \citet{Schirm2014} established a transition at $\jup \approx 4\mh 5$, with the emission from the cold phase dominating the small \jup regime. All our regions share a comparable distribution of cold ISM ($\lesssim 100 \K$) and have similar SLEDs at small $\jup$ (\fig{sled}), which confirms the conclusion of \citet{Schirm2014}.

In this regime, the behavior of our SLEDs corresponds to a ULIRG (or starburst) type of activity (as opposed to a disk, steady regime of star formation). However, at higher $\jup$, the arc deviates from the common ULIRG-like behavior and resembles more the high redshift disk. This is due to a different inter-cloud medium in the arc where the dense gas is highly concentrated into clouds surrounded by relatively diffuse medium ($\sim 10 \cc$, recall \fig{pdf}), like the massive clumps in high redshift gas rich disks \citep[see a discussion in][]{Bournaud2015}. Conversely, due to destroying tides, shear and the overall shorter separation between the star-forming regions (\sect{scale}), the inter-cloud media of the nuclei and the overlap gather larger amounts of dense and warm gas ($\sim 10^{2\mh 3} \cc$, $10^4 \K$), likely heated by stellar feedback which propagates further from the young stars in such inhomogeneous, porous media (see \citealt{Ohlin2019}, and the next section). Because these types of regions typically dominate the emission of (U)LIRGs (\fig{cocontrib}), they are good representatives of our large-scales SLEDs (``global'' and ``ULIRG'' in \fig{sled}).

The state of the medium surrounding the density peaks is critical to set the external pressure on the clouds, and possibly sets the mass of the star clusters formed \citep{Elmegreen1997, Ashman2001, Maji2017}. Although our model predict comparable pressures for the overlap and the nuclei (as estimated observationally by \citealt{Schirm2014}), a comparison on the mass of the clusters is likely inadequate when considering the extreme cases of nuclear clusters. However, our analysis also leads (at first order) to a higher pressure in the overlap than in the arc, that would translate into more massive clusters in the former region. Yet, comparable massive clusters are found in both regions in our simulation \citep{Renaud2015}, and in observations \citep[e.g.][]{Bastian2009}, which questions the sole role of pressure in setting the mass of young clusters.

\subsection{On the importance of capturing the structure of the cold ISM}
\label{sec:coldism}

Our results on the CO emission and the diversity of depletion times reflect the distribution of dense gas but also the nature of the surrounding, more diffuse ISM. The hierarchical nature of the ISM \citep{Elmegreen1997, Elmegreen2008} is here key to assess how star-forming seeds are fueled with gas, and how fluxes are emitted and absorbed. For instance, we have noted above that the decoupling of the stars from their gas nursery (mainly because of the dissipative nature of the ISM, \citealt{Renaud2013b}, or because of deformation of the clouds by tides or shear) can lead to significant drifts of young stars away from the densest regions of the clouds. In this context, the porosity of the ISM is equally important, particularly to regulate the propagation of feedback effects. This porosity is not only set by the level of turbulence (i.e. the Mach number), but also by the nature of turbulence (compressive vs. solenoidal). While a solenoidal-dominated turbulence tends to homogenize the ISM by smoothing out density contrasts (and thus slowing down or even quenching star formation), a compressive-dominated turbulence favors the formation of high density filaments and clumps, surrounded by low density cavities and chimneys. Such low density volumes then facilitate the propagation of feedback effects (ionization, energy, momentum, chemical enrichment) to large distances, but along preferred directions of least-resistance, as demonstrated by \citet{Ohlin2019}. It is thus likely that the regulation of star formation by feedback effects is also a function of the nature of turbulence, and thus, following the results presented here, the depletion time of the cloud itself.

To conclude, the role of turbulence is multiple: by setting the distribution of dense gas (at $\sim 10-100 \pc$ scale), it shapes the formation of the star-forming clouds (which can then fragment and gravitationally collapse \citealt{Vazquez2009}), it makes the low-density channels in which the propagation of feedback is facilitated, and it sets the optical depth through a series of high and low density structures. It is therefore of prime importance for models and simulations to capture the turbulence and its potential diversity over a large dynamical range.

\section{Conclusions}

We present an analysis of the spatial variations of the star formation activity in a numerical model of Antennae-like interacting galaxies at parsec resolution. We identify several mechanisms enhancing or slowing down star formation, and study their interplay and observable signatures in different regions of the galaxies. Our main conclusions are as follows.
\begin{itemize}
\item The interplay of the physical processes acting on star formation exhibits complex variations in space. Their timescales are comparable to that of the dynamics of interacting galaxies (i.e. a few $10 \Myr$), implying that the initial cause(s) for the formation (or destruction) of clouds can be long gone when star formation itself occurs. As a result, no clear (anti-)correlation can be found between causes (inflows, shocks, tidal and turbulent compression, shear, tidal stress) and consequences of star formation. Such correlations are, however, unambiguous when integrating over galactic scales and examining the time-evolution along the interaction and the merger \citep{Renaud2014b}.
\item All star-forming clouds yield enhanced SFRs and reduced depletion times with respect to isolated galaxies. At cloud-scale, a correlation exists between short depletion times ($\lesssim 40 \Myr$) and excess of dense gas visible as a secondary peak in the density PDF ($\gtrsim 10^4 \cc$). This occurs in all regions of the galaxies, with no apparent dependence on space (i.e. no gradient or symmetries). This translates at galactic scale into the entire system moving from the canonical Schmidt-Kennicutt relation of star-forming disks to the regime of starbursts, i.e. with a depletion times reduced by a factor $\approx 10$.
\item While all clouds have comparable sizes (radii of $\sim 30 \pc$ at their $50 \cc$ contour), the diversity of processes acting on the assembly and destruction of gaseous overdensities leads to different separations between the clouds in different regions of the galaxies ($120 \mh 270 \pc$). This has an imprint on the scale-dependence of measurements of the star formation activity like the depletion time.
\item The differences in the kpc-scale ISM between the arc and the other regions is visible in the CO emission, at high $J$ transitions. This is likely due to an excess of warm gas in the other regions, originating from the efficient propagation of stellar feedback in the inter-cloud medium ($\sim 10 \cc$). It follows that the arc shows CO SLEDs comparable to that of isolated disks (at low and high redshifts). The other regions contribute in larger proportions to the global emission of the system and thus imprint the CO emission of (U)LRIG-like galaxies.
\item We report significant spatial variations of the \aco factor across the galaxies, at least as important as the time variations along the course of the interaction i.e. between isolated progenitor phase and the peak of starburst. 
\item The excess of dense gas correlates with the onset of a compression-dominated turbulence, but the shape of the PDF fundamentally differs from that measured in simulations of isolated clouds. The reasons for this discrepancy are still under investigation. 
\end{itemize}

Although the links between the physical processes and their observational signatures are highly degenerated, combining different tracers and indicators of the amounts and distributions of dense gas (CO, HCN, HNC, etc.), its excitation (velocity dispersion, SLEDs including both low and high transitions), and the star formation activity (its rate and efficiency) would help disentangling comparable physical conditions. This could be particularly useful when individual star-forming regions cannot be resolved, e.g. in high redshift galaxies. For instance, at a given (surface) density of gas, a short depletion time traces an excess of dense gas, which calls for an kpc-scale trigger. In such a medium, an highly excited CO at $\jup \gtrsim 4\mh 5$ likely corresponds to shocks (or equivalently to an infall toward a unique convergence point). However, a more moderate SLED would indicate compression from larger scales, possibly of tidal and turbulent origin (like in the arc in our study). How these different states are combined and mix across the galaxy could then help inferring the underlying nature of the galaxy and its evolutionary stage in case of an interacting system. It would also provide constraints on the value of \aco to be adopted \citep[see also][]{Renaud2019}. Note however that such a diversity of mechanisms comes with a diversity of timescales, which can be of the same order as the dynamical time of galaxies, especially in fast evolving systems (e.g. interactions, bars, galactic centers). This blurs the identification of the physical processes at stake in a complex manner, which further calls for the use of as many independent tracers as available.

The identification of the physical trigger of the star formation enhancement in interacting galaxies is a difficult task, even in simulations. Our analysis reveals resemblances between the overlap and the two nuclei, with a comparable influence of shocks between the gas reservoirs and cloud-cloud collisions. This is balanced by a strong shear on the infalling material at the vicinity of the nuclei. In this respect, the arc stands out, as it hosts no cloud-cloud collisions, no shocks and no inflows. The only physical process left to explain the enhancement of star formation (high SFR and short depletion time) is that of tidal and turbulence compression \citep{Renaud2014b}. We note that clouds formed there are less numerous than in other regions, but once formed they have comparable properties (sizes, mass, PDF, depletion times) as in the most active regions. The similarities at $1\mh 30 \pc$ scales is also retrieved in the formation of massive star clusters \citep{Renaud2015}. The main difference we found is for the CO emission from the warm phase of the ISM ($\sim 10^4 \K$, $\jup \gtrsim 4\mh 5$). The key in determining observationally the physical trigger of enhanced star formation could thus be hidden in the medium surrounding the clouds themselves.

\begin{acknowledgements}
FR and OA acknowledge support from the Knut and Alice Wallenberg Foundation. OA acknowledges support from the Swedish Research Council (grant 2014-5791). This work was supported by GENCI (allocations A0030402192 and A0050402192) and PRACE (allocation pr86di) resources.
\end{acknowledgements}

\bibliographystyle{aa}

\begin{appendix}
\section{Comparison of CO SLEDs with observations}
\label{sec:obs}

\begin{figure}
\centering
\includegraphics{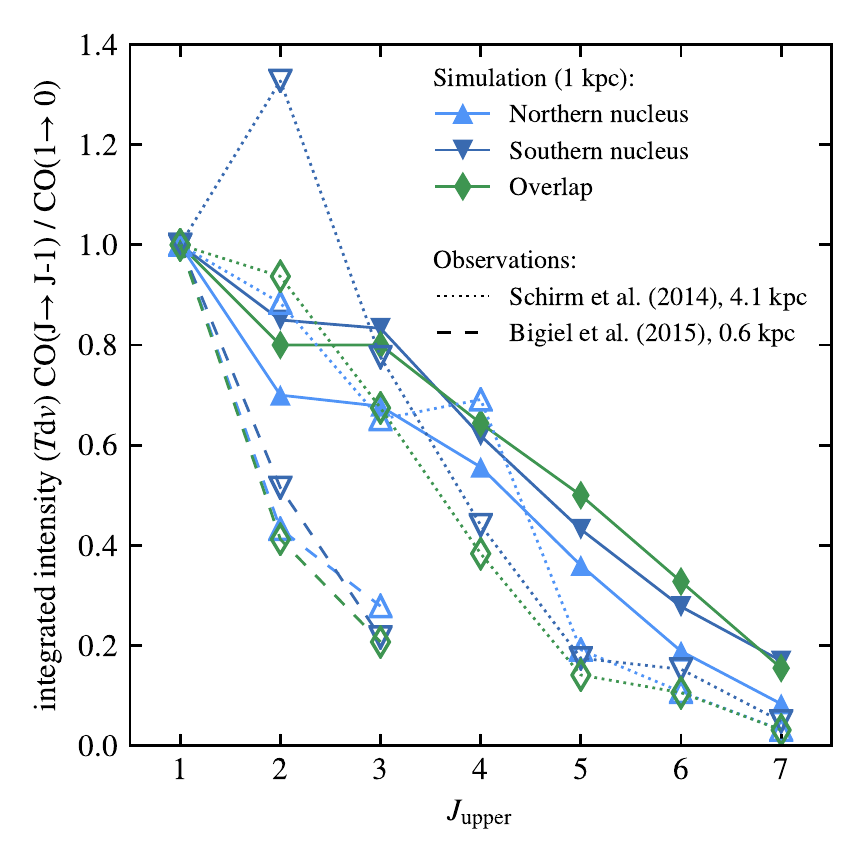}
\caption{SLEDs in velocity-integrated intensities from the simulation compared to observations from \citet{Schirm2014} and \citet{Bigiel2015} of the equivalent regions in the real Antennae. (We show here the average of the three overlap points presented in Bigiel et al.)}
\label{fig:obs}
\end{figure}

\fig{obs} compares the simulated SLEDs with that from observations in the two nuclei and the overlap. The data from \citet{Schirm2014} is obtained for a beam size of $43''$, while \citet{Bigiel2015} focuses on the emission peaks at a resolution of $6.5''$. Assuming a distance to the Antennae of $19 \mh 22 \Mpc$ \citep{Whitmore1999, Schweizer2008}, these correspond to $4.0 \mh 4.5 \kpc$ and $0.6 \mh 0.7 \kpc$ respectively.

We report a better agreement between our simulated SLEDs and the values of Schirm et al. than between the two observations, which likely originates from differences in beam sizes. While Bigiel et al. focused on the peaks of emission, the pointings in Schirm et al. cover a much larger area and thus encompass a larger fraction of gas with weaker emission. This is qualitatively closer to the approach followed in our simulations, where our 1 kpc regions gather a wide range of CO-emitting media.

When considering de-normalised values (not shown), our model yields intensities $\sim 11 \mh 18$ times higher than that of Schirm et al. (for $\jup \le 4$, see below), which is compatible with a scale factor of $\pi (4 \kpc /2)^2 / (1 \kpc)^2 \approx 13$ between our surface areas.

We note that our model yields stronger excitations at $\jup \gtrsim 5$, compared to Schirm et al. Following the arguments presented in \sect[s]{sled} and \ref{sec:ismstructure}, we interpret this difference as a dilution of the contribution from the warm gas ($\sim 1\mh 10 \cc$, $10^4 \K$) in the larger beams of Schirm, compared to our regions.

\end{appendix}

\end{document}